\documentclass[epj]{svjour}
\usepackage{graphics}
\begin{document}
\title{Truncation scheme of time-dependent density-matrix approach III}
\author{Mitsuru Tohyama\inst{1} \and Peter Schuck\inst{2,3}}
\institute{Kyorin University School of Medicine, Mitaka, Tokyo
181-8611, Japan \and Institut de Physique Nucl\'eaire, IN2P3-CNRS,
Universit\'e Paris-Sud, F-91406 Orsay Cedex, France \and
Laboratoire de Physique et de Mod\'elisation des Milieux Condens\'es, CNRS 
 et, Universit\'e Joseph Fourier, 25 Av. des Martyrs, BP 166, F-38042 
 Grenoble Cedex 9, France}
 \date{Received: date / Revised version: date}
\abstract{The time-dependent density-matrix theory (TDDM) where the Bogoliubov-Born-Green-Kirkwood-Yvon hierarchy
for reduced density matrices is truncated by approximating a three-body density matrix with one-body and two-body density matrices 
is applied to the Lipkin model.
It is shown that in the large $N$ limit the ground state in TDDM approaches the exact solution.
Various truncation schemes for the three-body density matrix are also tested for an extended three-level Lipkin model.
}
\maketitle
\section{Introduction}
The equations of motion for reduced density matrices have a coupling scheme known as 
the Bogoliubov-Born-Green-Kirkwood-Yvon (BBGKY) hierarchy \cite{Bonitz} where an $n$-body density matrix couples to
$n$-body and $n+1$-body density matrices. 
To solve the equations of motion for the one-body and two-body density matrices, we need to truncate the BBGKY hierarchy
at a two-body level. The simplest truncation scheme is to approximate  
a three-body density matrix with the antisymmetrized products of the one-body and two-body density matrices neglecting the correlated
part of the three-body density matrix (the three-body correlation matrix) \cite{WC,GT}. 
We refer to this truncation scheme as the time-dependent density-matrix theory (TDDM).
In some cases TDDM 
overestimates ground-state correlations \cite{TTS}, gives unphysical results \cite{schmitt,gherega} and causes instabilities of the obtained solutions \cite{toh12} 
for strongly interacting cases. Obviously the problems originate in the
truncation scheme where the three-body correlation matrix is completely neglected. This is related to the loss of $N$-representability \cite{coleman} which is preserved only in the case of the
untruncated BBGKY hierarchy.
To remedy to such difficulties of the naive truncation scheme,
better approximations for the three-body correlation matrix 
have been proposed. One is to approximate the three-body correlation matrix with the products of the 
correlated part of the two-body correlation matrices based on a perturbative consideration \cite{ts2014}, which corresponds to the cumulant expansion \cite{mazz1999}:
The three-body correlation matrix may also be called the three-body cumulant.
We refer to this truncation scheme as TDDM1. 
It was found in the applications of TDDM1 to the Lipkin model \cite{Lip} that TDDM1 can remedy the overestimation of ground-state correlations in TDDM.
It was also found that TDDM1 underestimates ground-state correlations in strongly interacting regions when the number of particles $N$ becomes large.
This suggests that higher-order effects which effectively reduce the contribution of the three-body correlation matrix should be taken into account.
We proposed another truncation scheme \cite{ts2017} which includes a normalization factor in the three-body correlation
matrix in TDDM1. We refer to it as TDDM2.
In this work we pursue our investigation of the cumulant approximation to the 
correlated part of the three-body density matrix which enters the equation of motion for the two-body density matrix. 
In particular we will continue in this sense our investigation of the two-level 
Lipkin model where we will find that TDDM solves this model exactly in the 
thermodynamic limit staying in the symmetry unbroken (spherical) basis.
Additionally we then also study the three-level Lipkin model \cite{li} which will turn 
out to have a quite different behavior with respect to the three-body correlations. 
We will find a particular approximation scheme for the cumulant expression of 
the three-body correlation matrix which also solves the three-level Lipkin model almost 
exactly for the ground state energy in the large $N$ limit staying in the 
symmetry unbroken description. However, this scheme will not be able to 
reproduce the rotational spectrum. A detailed discussion of this behavior will 
be given and the particularities of the three-level Lipkin model in this respect 
will be discussed.
At the end, we will try to give some general conclusions concerning the 
performance of the cumulant approximation to the three-body correlation matrix in 
the TDDM approach.

The paper is organized as follows. The three truncation schemes TDDM, TDDM1 and TDDM2 are explained in sect. 2. The results for the two and three-level Lipkin models are
presented in sect. 4 and sect. 4 is devoted to summary.

\section{Formulation}
We consider a system of $N$ fermions and assume that the Hamiltonian $H$ consisting of a one-body part and a two-body interaction
\begin{eqnarray}
H=\sum_{\alpha\alpha'}\langle\alpha|t|\alpha'\rangle a^\dag_\alpha a_{\alpha'}
+\frac{1}{2}\sum_{\alpha\beta\alpha'\beta'}\langle\alpha\beta|v|\alpha'\beta'\rangle
a^\dag_{\alpha}a^\dag_\beta a_{\beta'}a_{\alpha'},
\nonumber \\
\label{totalH}
\end{eqnarray}
where $a^\dag_\alpha$ and $a_\alpha$ are the creation and annihilation operators of a particle at
a single-particle state $\alpha$.
\subsection{Time-dependent density-matrix theories}
The TDDM approaches give the coupled equations of motion for the one-body density matrix (the occupation matrix) $n_{\alpha\alpha'}$
and the correlated part of the two-body density matrix $C_{\alpha\beta\alpha'\beta'}$.
These matrices are defined as
\begin{eqnarray}
n_{\alpha\alpha'}(t)&=&\langle\Phi(t)|a^\dag_{\alpha'} a_\alpha|\Phi(t)\rangle,
\\
C_{\alpha\beta\alpha'\beta'}(t)&=&\langle\Phi(t)|a^\dag_{\alpha'}a^\dag_{\beta'}
 a_{\beta}a_{\alpha}|\Phi(t)\rangle
\nonumber \\
&-&(n_{\alpha\alpha'}(t)n_{\beta\beta'}(t)-n_{\alpha\beta'}(t)n_{\beta\alpha'}(t)) ,
 \label{rho2}
\end{eqnarray}
where $|\Phi(t)\rangle$ is the time-dependent total wavefunction
$|\Phi(t)\rangle=\exp[-iHt] |\Phi(t=0)\rangle$.
We use units such that $\hbar=1$.
The equations of motion for $n_{\alpha\alpha'}$ and $C_{\alpha\beta\alpha'\beta'}$ are written as
\begin{eqnarray}
i \dot{n}_{\alpha\alpha'}&=&
\sum_{\lambda}(\epsilon_{\alpha\lambda}{n}_{\lambda\alpha'}-{n}_{\alpha\lambda}\epsilon_{\lambda\alpha'})
\nonumber \\
&+&\sum_{\lambda_1\lambda_2\lambda_3}
[\langle\alpha\lambda_1|v|\lambda_2\lambda_3\rangle C_{\lambda_2\lambda_3\alpha'\lambda_1}
\nonumber \\
&-&C_{\alpha\lambda_1\lambda_2\lambda_3}\langle\lambda_2\lambda_3|v|\alpha'\lambda_1\rangle],
\label{n1}
\end{eqnarray}
\begin{eqnarray}
i\dot{C}_{\alpha\beta\alpha'\beta'}&=&
\sum_{\lambda}(\epsilon_{\alpha\lambda}{C}_{\lambda\beta\alpha'\beta'}
+\epsilon_{\beta\lambda}{C}_{\alpha\lambda\alpha'\beta'}
\nonumber \\
&-&\epsilon_{\lambda\alpha'}{C}_{\alpha\beta\lambda\beta'}
-\epsilon_{\lambda\beta'}{C}_{\alpha\beta\alpha'\lambda})
\nonumber \\
&+&B_{\alpha\beta\alpha'\beta'}+P_{\alpha\beta\alpha'\beta'}+H_{\alpha\beta\alpha'\beta'}
\nonumber \\
&+&\sum_{\lambda_1\lambda_2\lambda_3}
[\langle\alpha\lambda_1|v|\lambda_2\lambda_3\rangle C_{\lambda_2\lambda_3\beta\alpha'\lambda_1\beta'}
\nonumber \\
&+&\langle\lambda_1\beta|v|\lambda_2\lambda_3\rangle C_{\lambda_2\lambda_3\alpha\alpha'\lambda_1\beta'}
\nonumber \\
&-&\langle\lambda_1\lambda_2|v|\alpha'\lambda_3\rangle C_{\alpha\lambda_3\beta\lambda_1\lambda_2\beta'}
\nonumber \\
&-&\langle\lambda_1\lambda_2|v|\lambda_3\beta'\rangle C_{\alpha\lambda_3\beta\lambda_1\lambda_2\alpha'}],
\label{r2}
\end{eqnarray}
where $C_{\alpha\beta\gamma\alpha'\beta'\gamma'}$ is the three-body correlation matrix
which is neglected in the original version of TDDM \cite{WC,GT}. 
The energy matrix $\epsilon_{\alpha\alpha'}$ is given by
\begin{eqnarray}
\epsilon_{\alpha\alpha'}=\langle \alpha|t|\alpha'\rangle
+\sum_{\lambda_1\lambda_2}
\langle\alpha\lambda_1|v|\alpha'\lambda_2\rangle_A 
n_{\lambda_2\lambda_1}
\label{hf},
\end{eqnarray}
where the subscript $A$ means that the corresponding matrix is antisymmetrized. 
The matrix $B_{\alpha\beta\alpha'\beta'}$ in eq. (\ref{r2}) does not contain $C_{\alpha\beta\alpha'\beta'}$ and describes 2p--2h and 2h--2p
excitations, where ${\rm p}$ and ${\rm h}$ refer to particle and hole states, respectively.
The terms $P_{\alpha\beta\alpha'\beta'}$ and $H_{\alpha\beta\alpha'\beta'}$
contain $C_{\alpha\beta\alpha'\beta'}$ and describe
p--p (and h--h) and p--h
correlations to infinite order, respectively \cite{GT}. These matrices are given in ref. \cite{GT}
but listed again in Appendix A for completeness.
Equations (\ref{n1}) and (\ref{r2}) satisfy the conservation laws of the total energy and the
total number of particles \cite{WC,GT}. 
As mentioned above,
$C_{\alpha\beta\gamma\alpha'\beta'\gamma'}$ in eq. (\ref{r2}) is neglected in TDDM \cite{WC,GT}.
In order to remedy difficulties of TDDM,
we have proposed a truncation scheme for $C_{\alpha\beta\gamma\alpha'\beta'\gamma'}$ \cite{ts2014} such that
\begin{eqnarray}
C_{\rm p_1p_2h_1p_3p_4h_2}&=&\sum_{\rm h}C_{\rm hh_1p_3p_4}C_{\rm p_1p_2h_2h},
\label{purt1}\\
C_{\rm p_1h_1h_2p_2h_3h_4}&=&\sum_{\rm p}C_{\rm h_1h_2p_2p}C_{\rm p_1ph_3h_4}.
\label{purt2}
\end{eqnarray}
These expressions were derived from perturbative consideration \cite{ts2014} using
the following CCD (Coupled-Cluster-Doubles)-like ground state wavefunction $|Z\rangle$ \cite{shavitt}
\begin{eqnarray}
|Z\rangle=e^Z|{\rm HF}\rangle
\end{eqnarray}
with 
\begin{eqnarray}
Z=\frac{1}{4}\sum_{\rm pp'hh'}z_{\rm pp'hh'}a^\dag_{\rm p}a^\dag_{\rm p'}a_{\rm h'}a_{\rm h},
\end{eqnarray}
where $|{\rm HF}\rangle$ is the Hartree-Fock (HF) ground state and  
$z_{\rm pp'hh'}$ is antisymmetric under the exchanges of ${\rm p} \leftrightarrow {\rm p'}$ and ${\rm h} \leftrightarrow {\rm h'}$.
We refer to the truncation scheme given by eqs. (\ref{purt1}) and (\ref{purt2}) as TDDM1. In the applications of TDDM1 to the Lipkin model
it was found that 
although TDDM1 effectively suppresses excess two-body correlations in small $N$ systems, 
it overestimates the contributions of the three-body correlation matrix in large $N$ systems.
To make TDDM1 applicable to large $N$ systems, we modified the truncation scheme \cite{ts2017} to  
\begin{eqnarray}
C_{\rm p_1p_2h_1p_3p_4h_2}&=&\frac{1}{\cal N}\sum_{\rm h}C_{\rm hh_1p_3p_4}C_{\rm p_1p_2h_2h},
\label{purt11}\\
C_{\rm p_1h_1h_2p_2h_3h_4}&=&\frac{1}{\cal N}\sum_{\rm p}C_{\rm h_1h_2p_2p}C_{\rm p_1ph_3h_4}.
\label{purt22}
\end{eqnarray}
The factor ${\cal N}$ is given by  
\begin{eqnarray}
{\cal N}&&=1+\frac{1}{4}\sum_{\rm pp'hh'}C_{\rm pp'hh'}C_{\rm hh'pp'},
\label{norm}
\end{eqnarray}
which plays a role in reducing the three-body correlation matrix in strongly interacting regions of the Lipkin model.
We refer to the truncation scheme given by eqs. (\ref{purt11}) and (\ref{purt22}) as TDDM2.
The conservation of the total energy and total particle number is not affected by the above approximations 
for the three-body correlation matrix because its anti-symmetry properties under the exchange of single-particle indices is respected.

\section{Results}
\subsection{Two-level Lipkin model}
First we consider the usual two-level Lipkin model.
The Lipkin model \cite{Lip} describes an $N$-fermions system with two
$N$-fold degenerate levels with energies $\epsilon/2$ and $-\epsilon/2$,
respectively. The upper and lower levels are labeled by quantum number
$p$ and $-p$, respectively, with $p=1,2,...,N$. We consider
the standard Hamiltonian
\begin{equation}
{H}=\epsilon {J}_{z}+\frac{V}{2}({J}_+^2+{J}_-^2),
\label{elipkin}
\end{equation}
where the operators are given as
\begin{eqnarray}
{J}_z&=&\frac{1}{2}\sum_{p=1}^N(c_p^{\dag}c_p-{c_{-p}}^{\dag}c_{-p}), \\
{J}_{+}&=&{J}_{-}^{\dag}=\sum_{p=1}^N c_p^{\dag}c_{-p}.
\end{eqnarray}
The HF ground state where the lowest single-particle states given by the operators $\{a_{-p}\}~~(p=1,2,\cdot\cdot\cdot N)$ are fully occupied
\begin{eqnarray}
|{\rm HF}\rangle=\prod_{p=1}^Na^\dag_{-p}|0\rangle
\end{eqnarray}
is obtained by the transformation
\begin{eqnarray}
\left(
\begin{array}{c}
a^\dag_{-p}\\
a^\dag_{p}
\end{array}
\right)=\left(
\begin{array}{cc}
\cos\alpha&\sin\alpha\\
-\sin\alpha&\cos\alpha
\end{array}
\right)
\left(
\begin{array}{c}
c^\dag_{-p}\\
c^\dag_{p}\\
\end{array}
\right).
\nonumber \\
\label{2l-trans}
\end{eqnarray}
The HF energy is given by
\begin{eqnarray}
E(\alpha)&=&-\frac{N\epsilon}{2}+N\epsilon \sin^2\alpha
\nonumber \\
&+&VN(N-1)\sin^2\alpha\cos^2\alpha.
\end{eqnarray}
Using $\chi=|V|(N-1)/\epsilon$, we can express $E_{\rm HF}$ which minimizes $E(\alpha)$ as 
\begin{eqnarray}
E_{\rm HF}=\left\{
\begin{array}{c}
-\frac{N\epsilon}{2}~~~~~(\chi\le 1)\\
\frac{N\epsilon}{4}(-\chi-1/\chi)~~~~~(\chi> 1),
\end{array}
\right .
\end{eqnarray}
where $\cos2\alpha=1/\chi$ is satisfied.

Since the solution in TDDM is closely related to that in the deformed Hartree-Fock theory (DHF) as shown below,
we discuss some properties of the DHF solution.
In DHF the occupation matrix is given by
\begin{eqnarray}
n_{-p}(n_{-p-p})&=&\cos^2\alpha,
\label{low}\\
n_{p}(n_{pp})&=&1-n_{-p}=\sin^2\alpha,
\label{up} \\
n_{p-p}&=&\cos\alpha\sin\alpha.
\end{eqnarray}
In DHF higher reduced density matrices have no correlated parts. For example the 2p--2h element of the two-body density matrix is expressed with
the occupation matrices as  
\begin{eqnarray}
\rho_{pp'-p-p'}&=&\langle{\rm DHF}|c^{\dag}_{-p}c^{\dag}_{-p'}c_{p'}c_p|{\rm DHF}\rangle
\nonumber \\
&=&n_{p-p}n_{p'-p'}=\cos^2\alpha\sin^2\alpha.
\label{2p2h}
\end{eqnarray}
Similarly, the ph--ph element is given by 
\begin{eqnarray}
\rho_{p-p'-pp'}(p\neq p')&=&\langle{\rm DHF}|c^{\dag}_{-p}c^{\dag}_{p'}c_{-p'}c_{p}|{\rm DHF}\rangle
\nonumber \\
&=&n_{p-p}n_{-p'p'}=\cos^2\alpha\sin^2\alpha.
\label{phph}
\end{eqnarray}
The three-body density matrix is also expressed as the antisymmetrized products of three occupation matrices
such that
\begin{eqnarray}
\rho_{p-p'p''pp'-p''}=\cos^2\alpha\sin^4\alpha,
\end{eqnarray}
which is the same as $n_{pp}\rho_{-p'p''p'-p''}$. This means that the correlated part of the three-body density matrix
$C_{p-p'p''pp'-p''}$ vanishes in DHF.

In TDDM (also in TDDM1 and TDDM2) the ground state is obtained using the adiabatic method \cite{toh2005,mazz2006} starting from the non-interacting "spherical" HF state
where the occupation matrix $n_{\alpha\alpha'}$ has no off-diagonal elements. Since the TDDM equations eqs. (\ref{n1}) and (\ref{r2})
maintain the symmetry, $n_{\alpha\alpha'}$ always has no off-diagonal elements. Similarly, $\rho_{pp'-p-p'}$ and $\rho_{p-p'-pp'}~(p\neq p')$ 
have no uncorrelated parts in TDDM. As shown below, there are strong similarities between the TDDM and DHF solutions such that
$C_{pp'-p-p'}$ and $C_{p-p'-pp'}$ correspond to eqs. (\ref{2p2h}) and 
(\ref{phph}), respectively. The diagonal elements eqs. (\ref{low}) and (\ref{up}) in DHF
also correspond to those in TDDM. This suggests the following relation for the expectation value of an operator $\hat{Q}$
\begin{eqnarray}
\langle \hat{Q}\rangle_{\rm TDDM}&\approx&\langle \Psi_{\rm DHF}|\hat{Q}|\Psi_{\rm DHF}\rangle
\nonumber \\
&\approx&\frac{1}{2} (\langle{\rm DHF}(\alpha)|\hat{Q}|{\rm DHF}(\alpha)\rangle
\nonumber \\
&+&\langle{\rm DHF}(-\alpha)|\hat{Q}|{\rm DHF}(-\alpha)\rangle),
\label{<q>}
\end{eqnarray}
where $|\Psi_{\rm DHF}\rangle$ is a 'spherical' wavefunction given by 
\begin{eqnarray}
|\Psi_{\rm DHF}\rangle=\frac{|{\rm DHF}(\alpha)\rangle+|{\rm DHF}(-\alpha)\rangle}{\sqrt{2}}.
\end{eqnarray}
Here $|{\rm DHF}(\alpha)\rangle$ means the deformed HF ground state with $\alpha$.  
In fact the following relations hold
\begin{eqnarray}
 \langle{\rm DHF}(\alpha)|{\rm DHF}(-\alpha)\rangle&=&(\cos^2\alpha-\sin^2\alpha)^N 
 \nonumber \\
 \\
 \langle{\rm DHF}(\alpha)|c^\dag_\alpha c_\beta|{\rm DHF}(-\alpha)\rangle&\propto& (\cos^2\alpha-\sin^2\alpha)^{N-1}
 \nonumber \\
 \\
 \langle{\rm DHF}(\alpha)|c^\dag_\alpha c^\dag_\beta c_{\beta'} c_{\alpha'}|{\rm DHF}(-\alpha)\rangle&\propto& (\cos^2\alpha-\sin^2\alpha)^{N-2}.
 \nonumber \\
\end{eqnarray}
Since $0<(\cos^2\alpha-\sin^2\alpha)<1$, the above expectation values between $|{\rm DHF}(\alpha)\rangle$ and $|{\rm DHF}(-\alpha)\rangle$ become
quite small for large $N$, justifying eq. (\ref{<q>}).
\begin{figure}
\resizebox{0.5\textwidth}{!}{%
\includegraphics{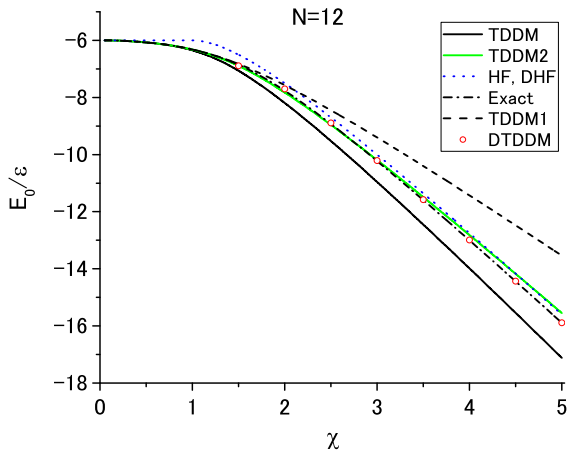}
}
\caption{Ground-state energy in TDDM (solid line)
as a function of $\chi=|V|(N-1)/\epsilon$ for $N=12$.
The results in TDDM1 where the three-body correlation matrix is given by eqs. (\ref{purt1}) and (\ref{purt2}) are shown with the dashed line.
The green (gray) line depicts the results in TDDM2 where the three-body correlation matrix is given by eqs. (\ref{purt11}) and (\ref{purt22}).
The results in HF and DHF($\chi>1$) are shown with the dotted line. The open circles (DTDDM) indicate the results in TDDM calculated using the 'deformed'
single-particle states and the DHF ground state.
The exact values are given by the dot-dashed line.}
\label{12e}
\end{figure}
\begin{figure}[h]
\resizebox{0.5\textwidth}{!}{
\includegraphics{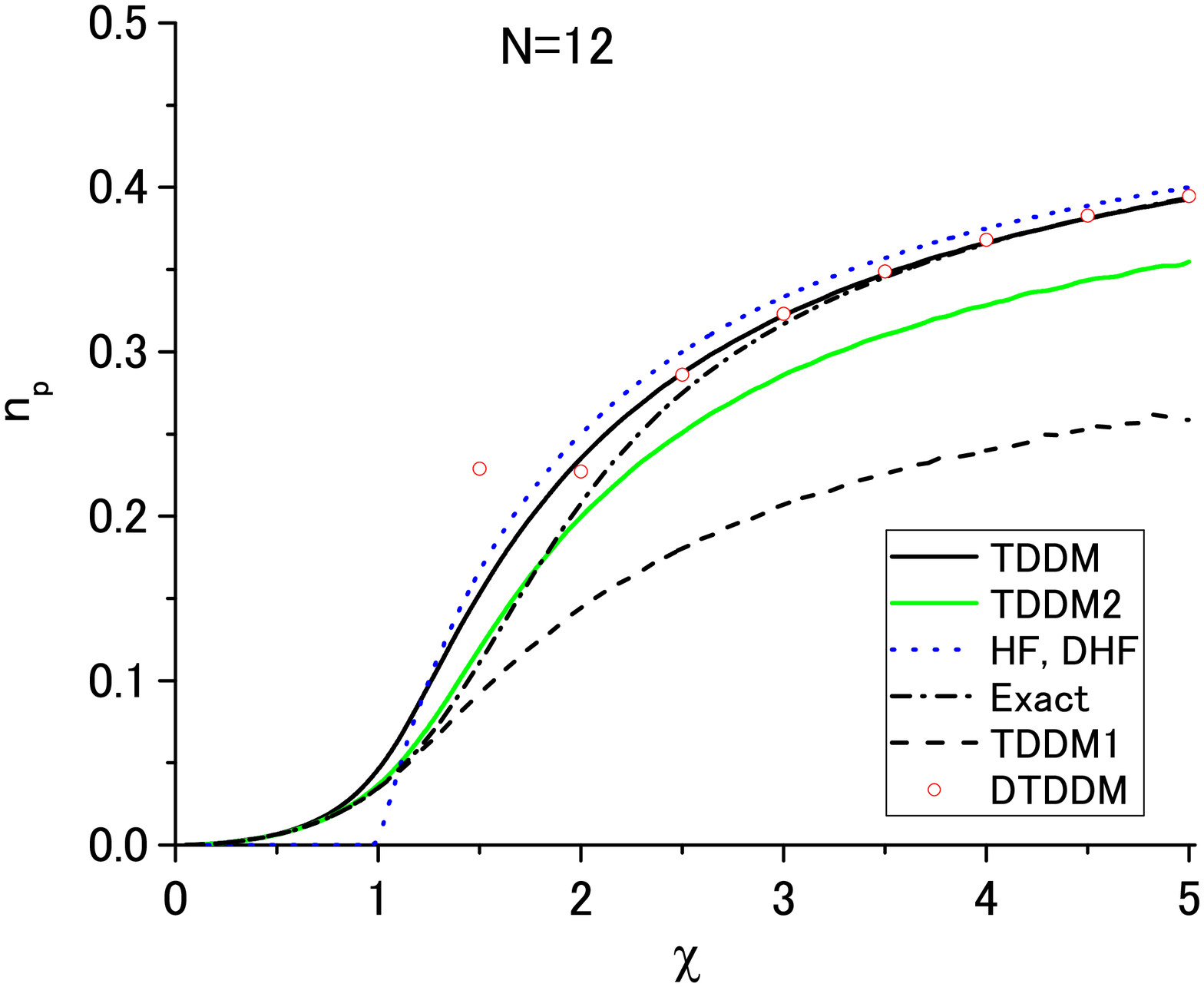}
}
\caption{Same as fig. \ref{12e} but for the occupation probability $n_{p}$ of the upper state.}
\label{12n}
\end{figure}

\begin{figure}[h]
\resizebox{0.5\textwidth}{!}{
\includegraphics{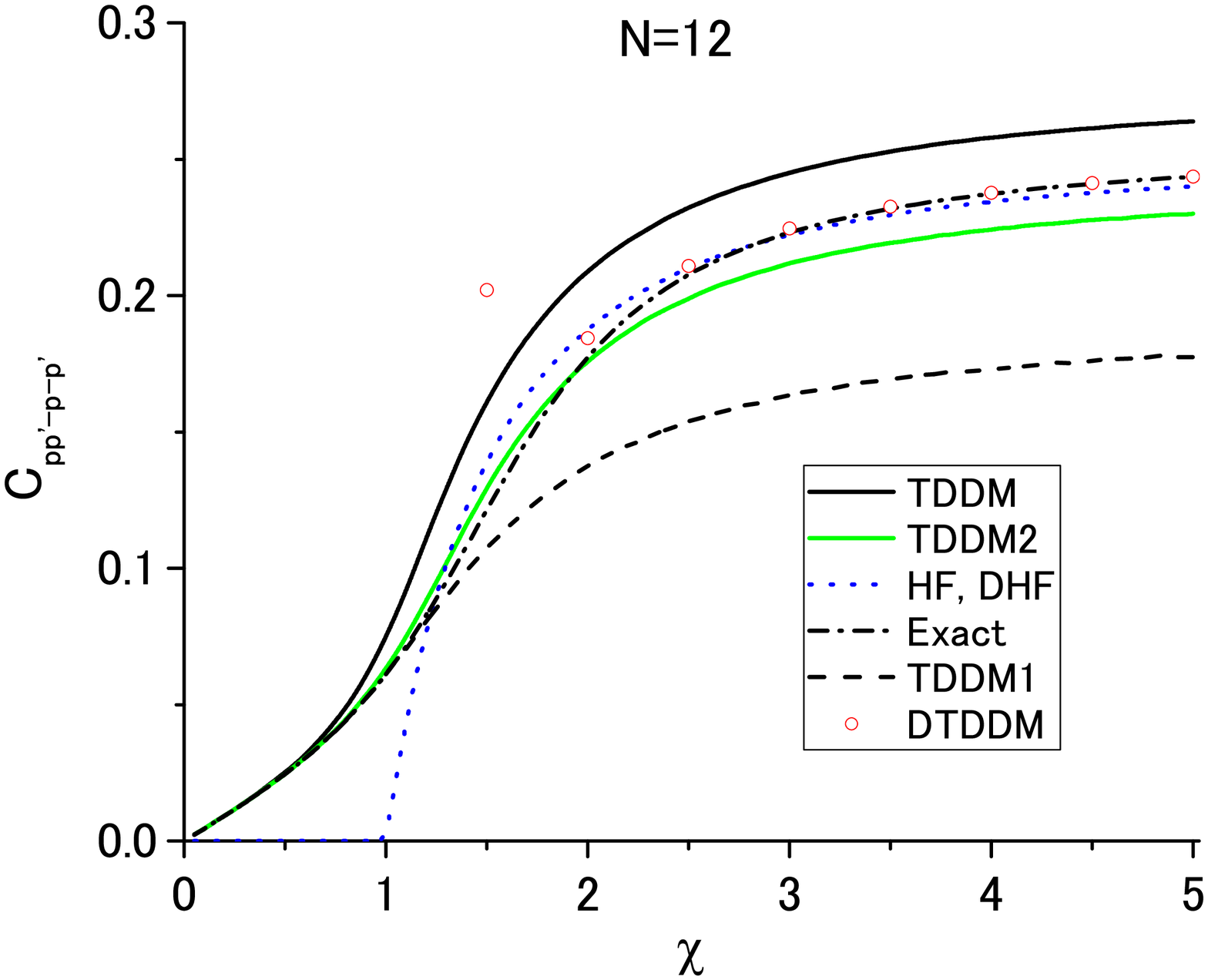}
}
\caption{Same as fig. \ref{12e} but for the 2p--2h element $C_{pp'-p-p'}$ of the two-body correlation matrix.
The results in DHF show $n_{p-p}^2$. }
\label{12c}
\end{figure}

\begin{figure}
\resizebox{0.5\textwidth}{!}{
\includegraphics{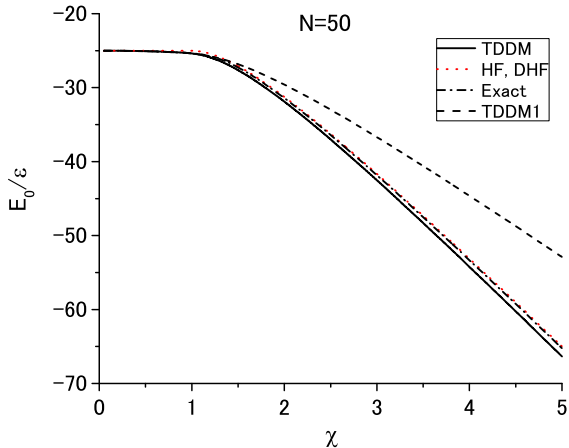}
}
\caption{Ground-state energy in TDDM (solid line)
as a function of $\chi$ for $N=50$.
The exact values are given by the dot-dashed line.
The dashed line depicts the results in TDDM1.
The results in TDDM2 lie between the TDDM results and the exact values and are not shown here.
The results in HF and DHF($\chi>1$) are shown with the dotted line but cannot be distinguished from the exact values 
in the scale of the figure except for the region $\chi\approx 1$.
}
\label{50e}
\end{figure}
\begin{figure}[h]
\resizebox{0.5\textwidth}{!}{
\includegraphics{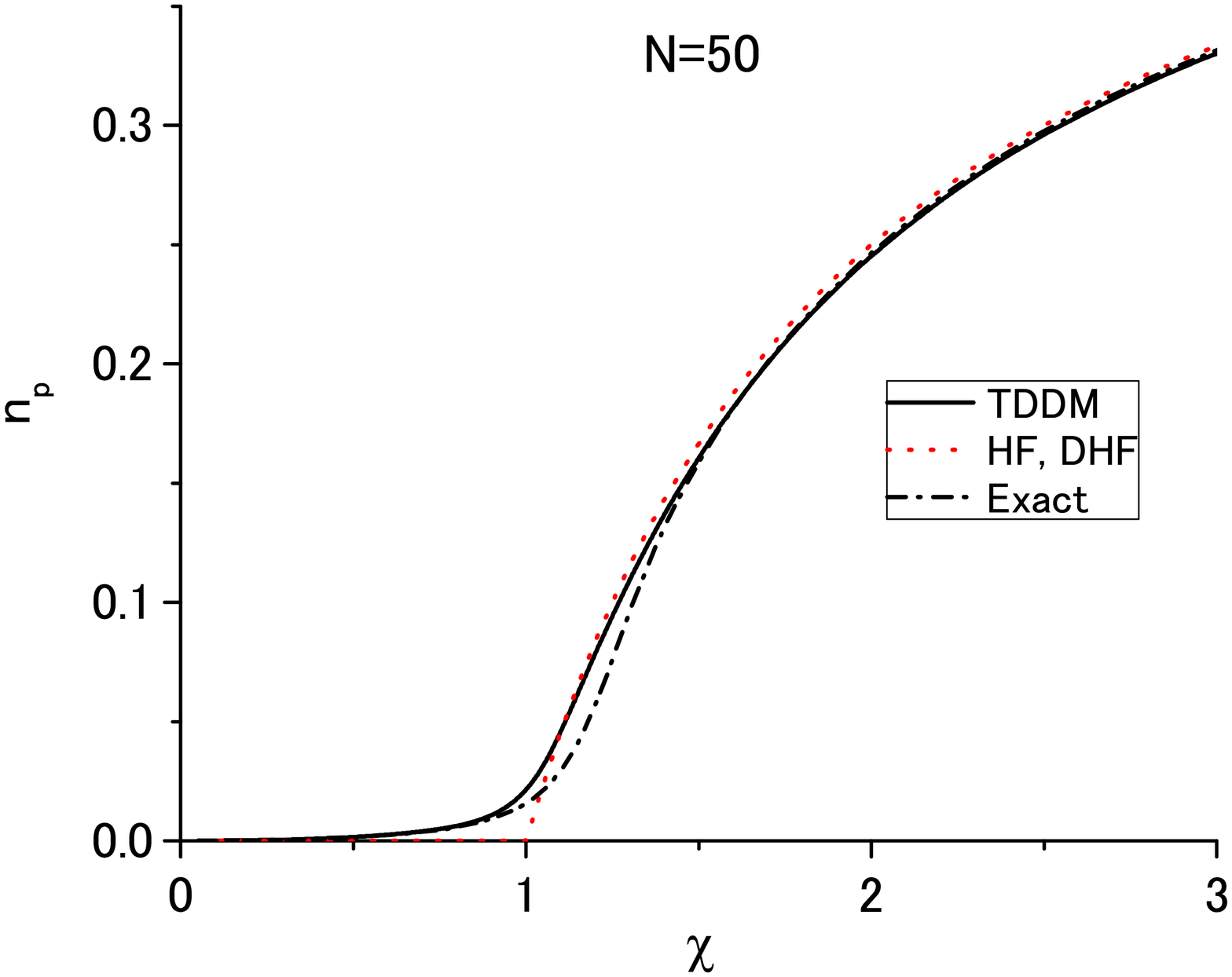}
}
\caption{Same as fig. \ref{50e} but for the occupation probability $n_{p}$ of the upper state.
}
\label{50n}
\end{figure}

\begin{figure}[h]
\resizebox{0.5\textwidth}{!}{
\includegraphics{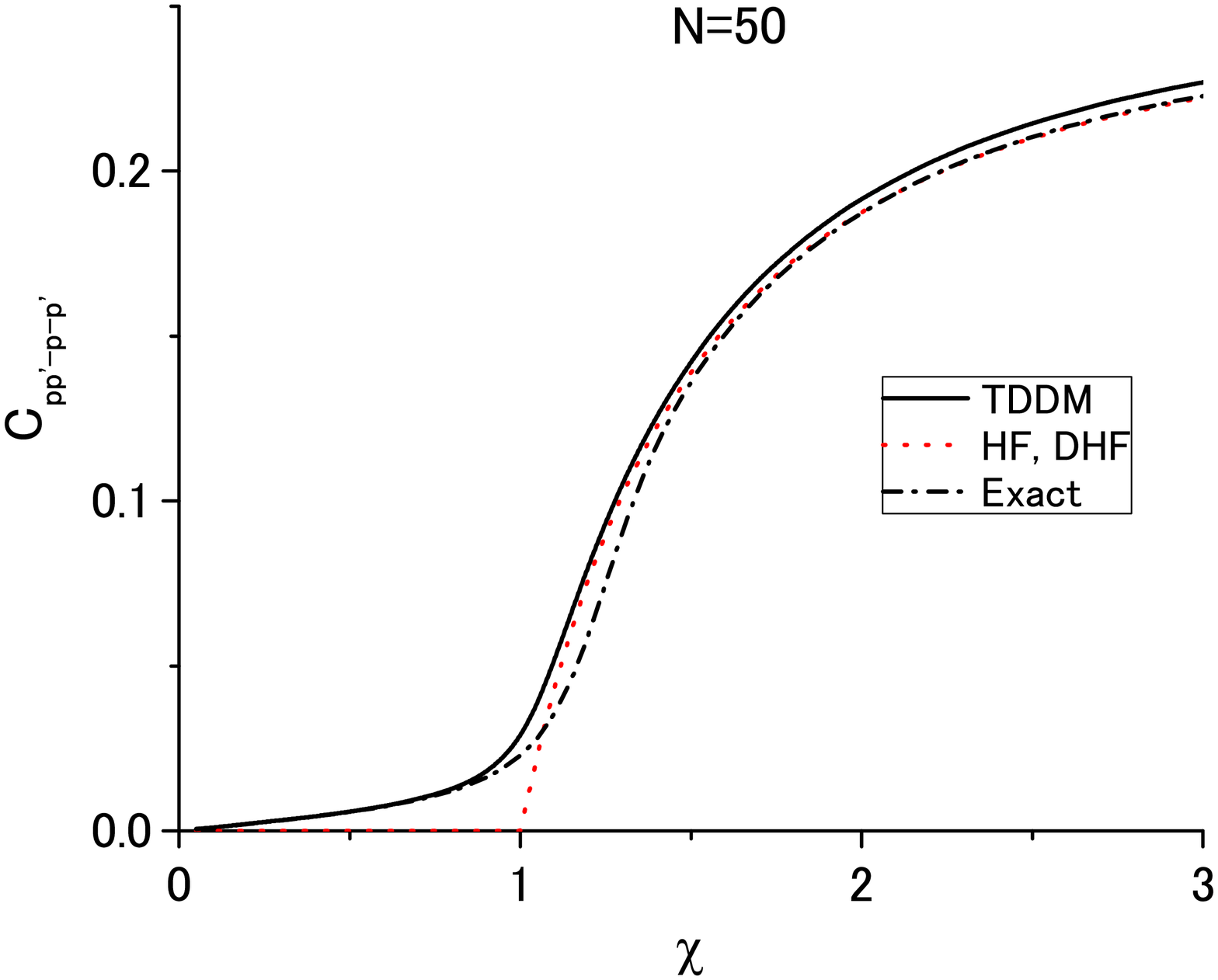}
}
\caption{Same as fig. \ref{50e} but for the 2p--2h element $C_{pp'-p-p'}$ of the two-body correlation matrix.
The results in DHF show $n_{p-p}^2$. }
\label{50c}
\end{figure}

\begin{figure}[h]
\resizebox{0.5\textwidth}{!}{
\includegraphics{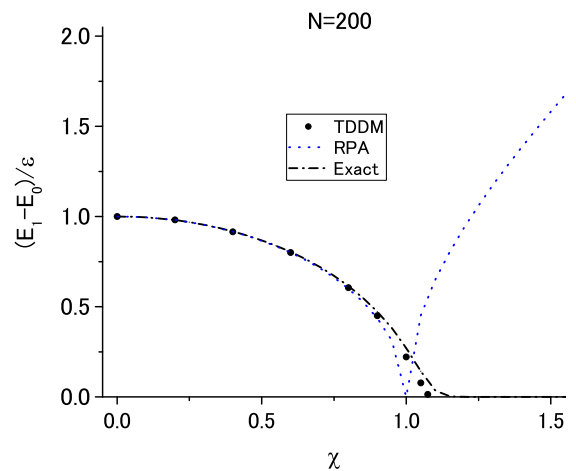}
}
\caption{Excitation energy of the first excited state calculated in TDDM (dots) a function of $\chi$ for $N=200$. The exact values are shown with the dot-dashed line.
The dotted line depicts the results in RPA.}
\label{200ex1}
\end{figure}

\begin{figure}[h]
\resizebox{0.5\textwidth}{!}{
\includegraphics{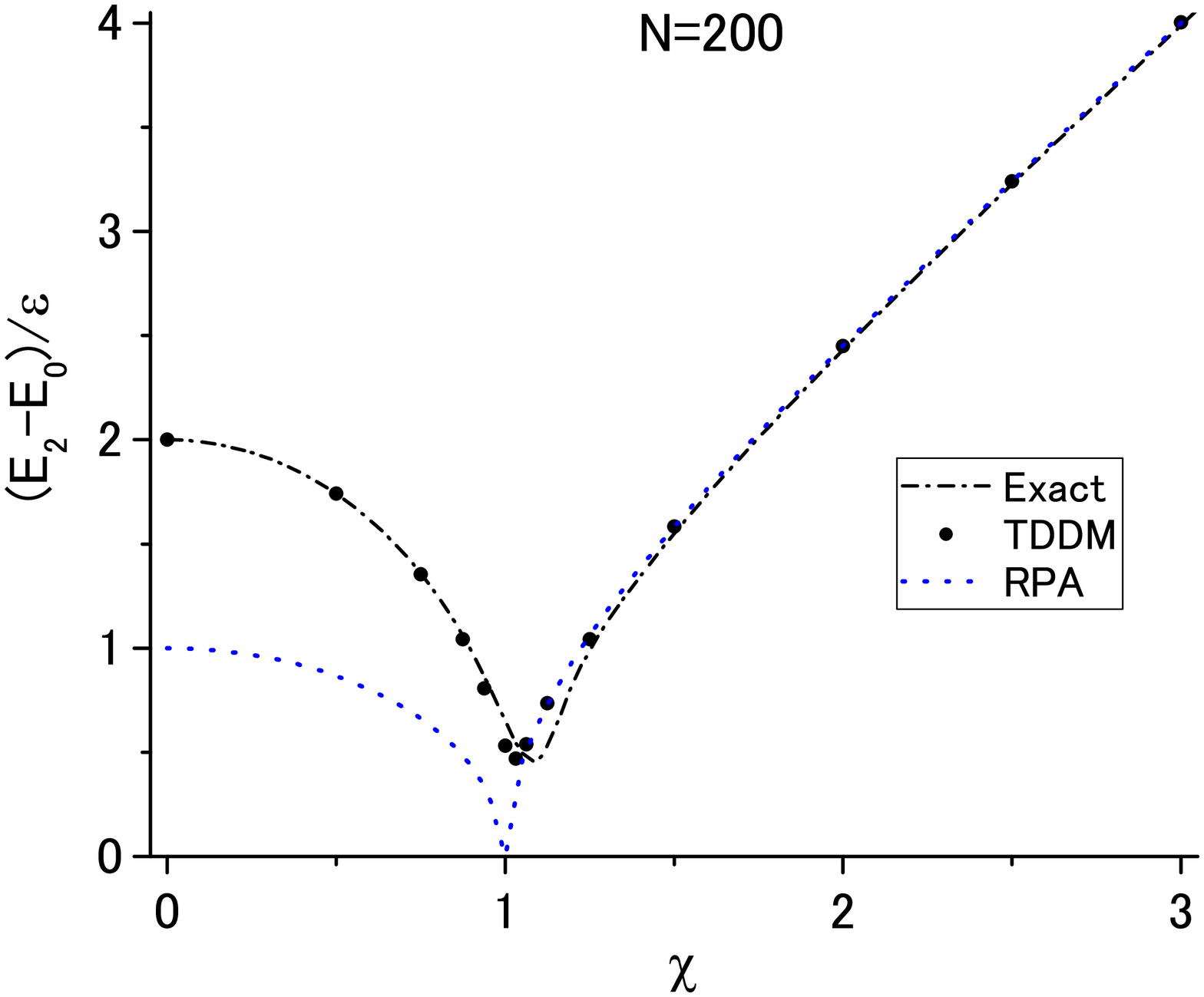}
}
\caption{Same as fig. \ref{200ex1} but for the second excited state. The dotted line depicts the RPA results for the first excited state.}
\label{200ex2}
\end{figure}
To illustrate how TDDM, TDDM1 and TDDM2 behave in small $N$ regions, we first present the results for $N=12$.
The ground-state energy $E_0$ calculated in TDDM (solid line) is shown in fig. \ref{12e} as a function of $\chi$.
The results in TDDM1 where the three-body correlation matrix is given by eqs. (\ref{purt1}) and (\ref{purt2}) are shown with the dashed line.
The results in TDDM2 where the three-body correlation matrix is given by eqs. (\ref{purt11}) and (\ref{purt22}) are shown with the green (gray) line.
The results in HF and DHF ($\chi>1$) and the exact values are shown with the dotted and dot-dashed lines, respectively.
The open circles indicate the results in TDDM calculated using the 'deformed' (symmetry broken)
single-particle states and the DHF ground state as the starting ground state. We refer to this scheme as the deformed TDDM (DTDDM). When the 'deformed' basis is used,
the two-body correlation matrix is so small that both DTDDM and the deformed TDDM1 give similar results.
The exact values are given by the dot-dashed line.
In this small $N$ system
TDDM overestimates two-body correlations \cite{ts2014,ts2017} and TDDM1 underestimates them in strongly interacting regions. TDDM2 cures this problem and 
the agreement with the exact solutions is much improved.  DTDDM also provides the additional correlation energy missing in DHF.
The occupation probability $n_{p}$ of the upper state and the 2p--2h element $C_{pp'-p-p'}$ of the correlation matrix calculated in TDDM 
are shown in figs. \ref{12n} and \ref{12c}, respectively. In DHF $n_{p-p}^2$ is shown as a quantity corresponding to $C_{pp'-p-p'}$.
TDDM1 and TDDM2 describe well $E_0$, $n_{p}$ and $C_{pp'-p-p'}$ in the transition region $\chi\approx 1$, where DHF fails to reproduce the exact values. 
Figures \ref{12n} and \ref{12c} indicate that the improvement in $E_0$ from TDDM to TDDM2 is due to the appropriate reduction of $C_{pp'-p-p'}$.
Figures \ref{12e}--\ref{12c} also show that DHF becomes good approximation with increasing interaction strength and that the improvement in $E_0$ from DHF to DTDDM 
is brought by that in $n_p$. The deviation of $n_p$ and $C_{pp'-p-p'}$ in DTDDM at $\chi=1.5$ is due to the fact that the deformed state becomes unstable in DTDDM near $\chi=1$. 

Now we present the TDDM results for $N=50$.
The ground-state energy $E_0$ calculated in TDDM (solid line) is shown in fig. \ref{50e} as a function of $\chi$.
The results in TDDM1 are given by the dashed line.
The dotted and dot-dashed lines depict the results in HF and DHF ($\chi>1$) and the exact values, respectively.
The results in HF and DHF cannot be distinguished from the exact values in the scale of the figure except for the region $\chi\approx 1$.
The results in TDDM2 lie between the TDDM results and the exact values and are not shown here.
In this large $N$ system, the ground-state energies in TDDM and TDDM2 agree well with the exact solutions and the DHF energies also become close to the exact values including the transition region. 
This is also seen in $n_p$ and $C_{pp'-p-p'}$ shown in figs. \ref{50n} and \ref{50c}, respectively.
The factor ${\cal N}$ in eqs. (\ref{purt11}) and (\ref{purt22}) plays a role in drastically reducing the three-body correlation matrix,
which makes TDDM2 almost equivalent to TDDM \cite{ts2017}. In DHF the three-body density matrix has no correlated parts. The fact that DHF becomes good approximation for the ground
state of the Lipkin model with
increasing number of particles explains that TDDM which has no three-body correlation matrix becomes good approximation for large $N$.
We checked that TDDM gives the almost exact ground state for much larger $N$. 

Finally we present the excitation energies of the first and second excited states calculated in TDDM for $N=200$ where TDDM gives the almost exact ground-state.
The energy of the first excited state is calculated using the TDDM equations eqs. (\ref{n1}) and (\ref{r2}) and assuming $n_{p-p}$, $C_{pp'-pp'}$ and 
$C_{-p-p'-pp'}$ are small. The second excited state which has the same symmetry as the ground state is calculated in a different manner. 
In the adiabatic method used for the ground-state calculations there is a small mixing of excited states which have the same symmetry as
the ground state. The excitation energy of the second excited state can be estimated from the frequency of the small oscillation
of $n_{pp}$ in the adiabatic method. The obtained results for the first and second excited states are compared with the exact solutions (dot-dashed line) in figs. \ref{200ex1}
and \ref{200ex2}, respectively. The dots in figs. \ref{200ex1} and \ref{200ex2} depict the TDDM results. 
The results in RPA for the first excited state are shown in figs. \ref{200ex1} and \ref{200ex2} with the dotted line. In contrast to RPA 
TDDM describes the decreasing excitation energy of the first excited state with increasing interaction strength beyond $\chi=1$. For $\chi>1.073$ TDDM does not give an oscillating solution to the first excited state.
This indicates that the three-body correlation matrix plays some role in the first excited state.
With increasing $\chi$ the RPA results for the first excited state approach the exact results for the second excited states as shown in fig. \ref{200ex2}.
The agreement of the TDDM results with the exact values is good for both the first and second excited states. This is related to the fact that 
TDDM becomes exact with increasing number of particles.

In this subsection we found out that in the two-level Lipkin model the three-body 
correlation matrix vanishes in the large $N$ (thermodynamic) limit and that TDDM solves the ground state of
this model exactly staying in the symmetry unbroken description. How far our 
findings can be transposed to other models of this type of a one site spin 
model is an open question but our studies may help to analyze the situation 
also in other cases.
We now will turn to the case of the three-level Lipkin model.

\subsection{Three-level Lipkin model}
Next we consider the three-level Lipkin model \cite{li} with the single-particle levels labeled 0, 1 and 2.
Since the number of unoccupied states is larger than that of the occupied states, the three-level Lipkin model may be more realistic than the
two-level Lipkin model and it has often been used to test extended mean-field theories \cite{hol,hagino,delion}.  We choose the Hamiltonian which is invariant under the exchange of 1 and 2:
\begin{eqnarray}
H=\epsilon(\hat{n}_1+\hat{n}_2)+\frac{V}{2}\left(K_1^2+K_2^2+(K_1^\dag)^2+(K_2^\dag)^2\right),
\label{3lH}
\nonumber \\
\end{eqnarray}
where 
\begin{eqnarray}
\hat{n}_\alpha=\sum_{i=1}^N c^\dag_{\alpha i}c_{\alpha i}~~~~~~~~~~\alpha=~0,~1,~2,
\label{spq}
\\
K_\alpha=\sum_{i=1}^N c^\dag_{\alpha i}c_{0i}~~~~~~~~~~\alpha=~1,~2.
\end{eqnarray}
The HF ground state where the lowest single-particle states given by the operators $\{a_{0i}\}~~(i=1,2,\cdot\cdot\cdot N)$ are fully occupied
\begin{eqnarray}
|{\rm HF}\rangle=\prod_{i=1}^Na^\dag_{0i}|0\rangle
\end{eqnarray}
is obtained by the transformation
\begin{eqnarray}
\left(
\begin{array}{c}
a^\dag_{0i}\\
a^\dag_{1i}\\
a^\dag_{2i}
\end{array}
\right)=\left(
\begin{array}{ccc}
\cos\alpha&\cos\beta\sin\alpha&\sin\beta\sin\alpha\\
-\sin\alpha&\cos\beta\cos\alpha&\sin\beta\cos\alpha\\
0&-\sin\beta&\cos\beta
\end{array}
\right)
\left(
\begin{array}{c}
c^\dag_{0i}\\
c^\dag_{1i}\\
c^\dag_{2i}
\end{array}
\right).
\nonumber \\
\label{3l-trans}
\end{eqnarray}
The HF energy is independent of $\beta$ and given by
\begin{eqnarray}
E(\alpha,\beta)=N\epsilon \sin^2\alpha+VN(N-1)\sin^2\alpha\cos^2\alpha.
\end{eqnarray}
Using $\chi=|V|(N-1)/\epsilon$, we can express $E_{\rm HF}$ which minimizes $E(\alpha,\beta)$ as 
\begin{eqnarray}
E_{\rm HF}=\left\{
\begin{array}{c}
0~~~~~(\chi\le 1)\\
\frac{N\epsilon}{4}(2-\chi-1/\chi)~~~~~(\chi> 1).
\end{array}
\right .
\end{eqnarray}
The ground-state wavefunction which has the symmetry under the exchange of 1 and 2 may be written as 
\begin{eqnarray}
|\Psi\rangle=\frac{1}{\pi}\int_{-\pi/2}^{\pi/2}|{\rm DHF}(\alpha,\beta)\rangle d\beta,
\label{total}
\end{eqnarray}
where $|{\rm DHF}(\alpha,\beta)\rangle$ is the DHF ground state with $\alpha$ and $\beta$. In the following we show that eq. (\ref{total})
gives a non-vanishing three-body correlation matrix.
To evaluate an expectation value of an operator $\hat{Q}$ using eq. (\ref{total}), we need the overlap $\langle{\rm DHF}(\alpha,\beta')|\hat{Q}|{\rm DHF}(\alpha,\beta)\rangle$. 
We show some examples of the overlaps using $X=\cos^2\alpha+\cos(\beta'-\beta)\sin^2\alpha$ \cite{hol}:
\begin{eqnarray}
&\langle{\rm DHF}&(\alpha,\beta')|{\rm DHF}(\alpha,\beta)\rangle=X^N,\\
&\langle{\rm DHF}&(\alpha,\beta')|c^\dag_{11}c_{11}|{\rm DHF}(\alpha,\beta)\rangle
\nonumber \\
&=&\cos\beta'\cos\beta\sin^2\alpha X^{(N-1)},\\
&\langle{\rm DHF}&(\alpha,\beta')|c^\dag_{01}c^\dag_{02}c_{12}c_{11}|{\rm DHF}(\alpha,\beta)\rangle
\nonumber \\
&=&\cos^2\beta\cos^2\alpha\sin^2\alpha X^{(N-2)},\\
&\langle{\rm DHF}&(\alpha,\beta')|c^\dag_{01}c^\dag_{12}c_{02}c_{11}|{\rm DHF}(\alpha,\beta)\rangle
\nonumber \\
&=&\cos\beta'\cos\beta\cos^2\alpha\sin^2\alpha X^{(N-2)},\\
&\langle{\rm DHF}&(\alpha,\beta')|c^\dag_{11}c^\dag_{12}c^\dag_{03}c_{13}c_{02}c_{11}|{\rm DHF}(\alpha,\beta)\rangle
\nonumber \\
&=&\cos^2\beta'\cos^2\beta\cos^2\alpha\sin^4\alpha X^{(N-3)},
\end{eqnarray}
where $01,~11,~\cdot\cdot\cdot$ mean the quantum number $(\alpha,i)$ for the single-particle state given in eq. (\ref{spq}). All components of the density matrices
do not depend on the quantum number $i$.
Since $X^N$ for large $N$ is sharply peaked at $\beta'-\beta\approx 0$, the diagonal assumption $\beta'=\beta$ is justified for the overlaps. 
Then the density matrices obtained with eq. (\ref{total}) are given as follows after the $\beta$ integration 
(we choose magnetic quantum numbers so that the expressions are general but in the end there will, of course, be no dependence on magnetic states):
\begin{eqnarray}
n_{\rm pp}&=&\langle\Psi|c^\dag_{11}c_{11}|\Psi\rangle=\frac{1}{2}\sin^2\alpha,
\label{npp}
\\
C_{\rm pp'hh'}&=&\langle\Psi|c^\dag_{01}c^\dag_{02}c_{12}c_{11}|\Psi\rangle
\nonumber \\
&=&\frac{1}{2}\cos^2\alpha\sin^2\alpha,
\label{pphh}
\\
C_{\rm ph'hp'}&=&\langle\Psi|c\dag_{01}c\dag_{12}c_{02}c_{11}|\Psi\rangle
\nonumber \\
&=&\frac{1}{2}\cos^2\alpha\sin^2\alpha,
\label{cphph}
\\
\rho_{\rm ph'p''pp'h''}&=&
\langle\Psi|c^\dag_{11}c^\dag_{12}c^\dag_{03}c_{13}c_{02}c_{11}|\Psi\rangle
\nonumber \\
&=&\frac{3}{8}\cos^2\alpha\sin^4\alpha.
\end{eqnarray}
Since $\rho_{\rm ph'p''pp'h''}$ contains four particle-state indices, the $\beta$ integration makes it impossible to express $\rho_{\rm ph'p''pp'h''}$ 
by the product of $n_{\rm pp}$ and $C_{\rm ph'hp'}$ which respectively have two particle-state indices.
From the above expressions we obtain the correlated part of the three-body density matrix as
\begin{eqnarray}
C_{\rm ph'p''pp'h''}&=&C_{11,02,13,11,12,03}
\nonumber \\
&=&\langle\Psi|c^\dag_{11}c^\dag_{12}c^\dag_{03}c_{13}c_{02}c_{11}|\Psi\rangle
\nonumber \\
&-&\langle\Psi|c^\dag_{11}c_{11}|\Psi\rangle\langle\Psi|c^\dag_{12}c^\dag_{03}c_{13}c_{02}|\Psi\rangle
\nonumber \\
&=&\frac{1}{8}\cos^2\alpha\sin^4\alpha.
\label{c3dhf}
\end{eqnarray}
Similarly, the three-body correlation matrix $C_{\rm ph'h''p'hh''}=C_{11,02,03,12,01,03}$ corresponding to eq. (\ref{purt2}) is
evaluated using the DHF wavefunction eq. (\ref{total}) and it is found that $C_{11,02,03,12,01,03}=0$
as is the case of the two-level Lipkin model. This is because $\rho_{\rm ph'h''p'hh''}$ contains only two particle-state indices, which allows us to express it as $n_{\rm hh}C_{\rm ph'hp'}$.
On the other hand, the three-body correlation matrix of the form 
$C_{\rm pp'p''hp'h''}$,
which enters the equation of motion for $C_{\rm ph'p'h}$, is also non-vanishing because it contains four particle-state indices. Using eq. (\ref{total}) it is 
evaluated to be the same as eq. (\ref{c3dhf}) that is ($\cos^2\alpha\sin^4\alpha/8$).
The above analysis indicates that the three-body correlation matrix in the strongly interacting region $(\chi>1)$ of the three-level Lipkin model is quite different from that of the two-level Lipkin model
and that it should be properly treated. However, 
extending the truncation scheme to include the equation of motion for the three-body correlation matrix is a difficult task and is not pursued in this work.
In the following we just point out
that there exists a simple truncation scheme for the three-body correlation matrix which gives a good description of the ground state
for the whole range of the interaction strength. In this approach the quadratic approximation for the pph--pph type three-body correlation matrix given by eq. (\ref{purt1}) is used
and the phh--phh type given by eq. (\ref{purt2}) is omitted considering the above analysis based on the DHF wavefunction.
We refer to this truncation scheme as TDDM1-b. We first present the results in TDDM1-b and then discuss why TDDM1-b works.


\begin{figure}
\resizebox{0.5\textwidth}{!}{
\includegraphics{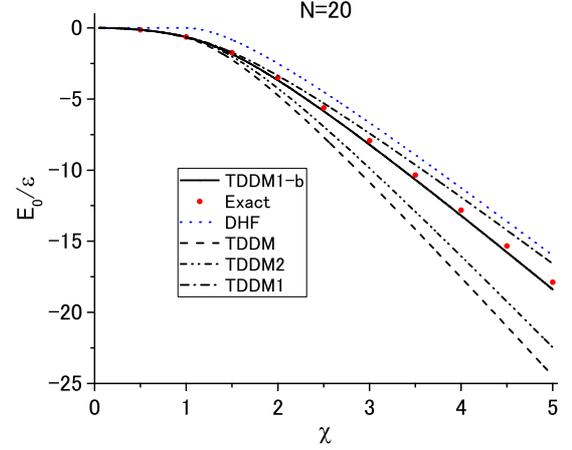}
}
\caption{Ground-state energy of the three-level Lipkin model calculated in TDDM1-b (solid line)
as a function of $\chi$ for $N=20$. The dashed, dot-dashed and two-dot-dashed lines depict the results in TDDM, TDDM1 and TDDM2, respectively.
The results in HF and DHF($\chi>1$) are shown with the dotted line.
The exact values are given by the dots.}
\label{20e3l}
\end{figure}
\begin{figure}
\resizebox{0.5\textwidth}{!}{
\includegraphics{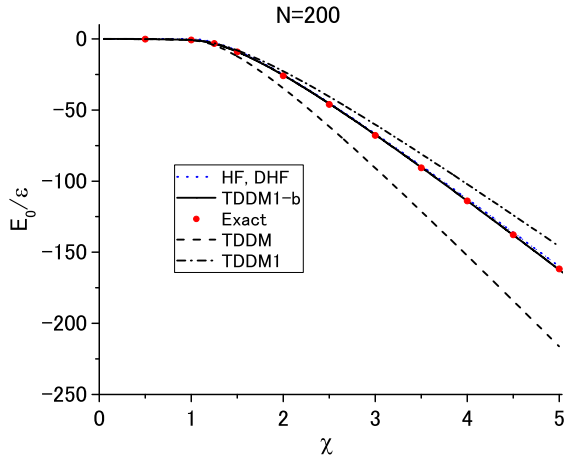}
}
\caption{Ground-state energy of the three-level Lipkin model calculated in TDDM1-b (solid line)
as a function of $\chi$ for $N=200$. The exact values are given by the dots.
The dashed and dot-dashed lines depict the results in TDDM and TDDM1, respectively.
The results in HF and DHF($\chi>1$) are shown with the dotted line but cannot be easily distinguished from the TDDM1-b results in the scale of the figure.
}
\label{200e3l}
\end{figure}

\begin{figure}
\resizebox{0.5\textwidth}{!}{
\includegraphics{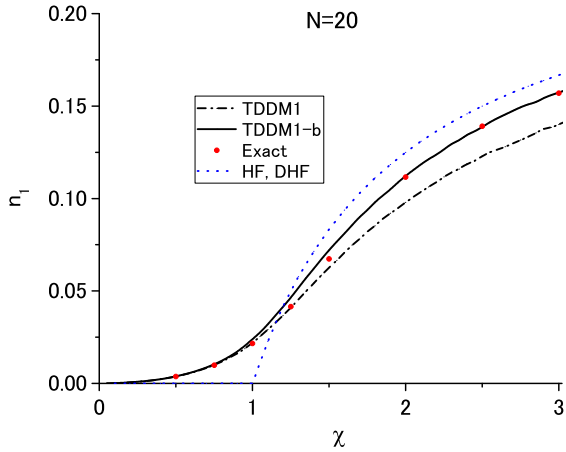}
}
\caption{Occupation probability of the state 1 of the three-level Lipkin model calculated in TDDM1-b (solid line)
as a function of $\chi$ for $N=20$. The dot-dashed line depicts the results in TDDM1.
The results in HF and DHF($\chi>1$) given by eq. (\ref{npp}) are shown with the dotted line.
The exact values are given by the dots.}
\label{20n3l}
\end{figure}

\begin{figure}
\resizebox{0.5\textwidth}{!}{
\includegraphics{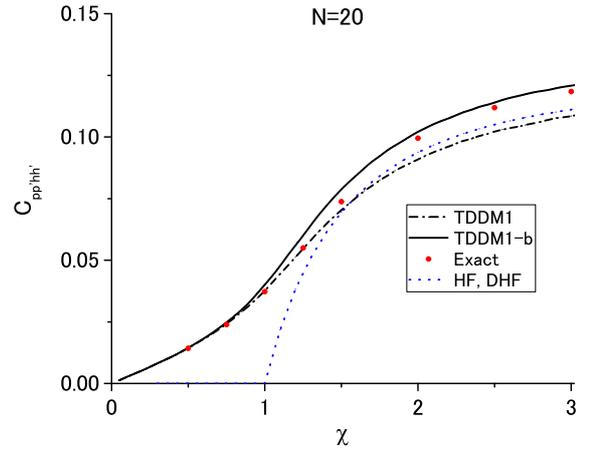}
}
\caption{Same as fig. \ref{20n3l} but for $C_{\rm pp'hh'}$. The results in HF and DHF are given by eq. (\ref{pphh}).}
\label{20c3l}
\end{figure}

\begin{figure}
\resizebox{0.5\textwidth}{!}{
\includegraphics{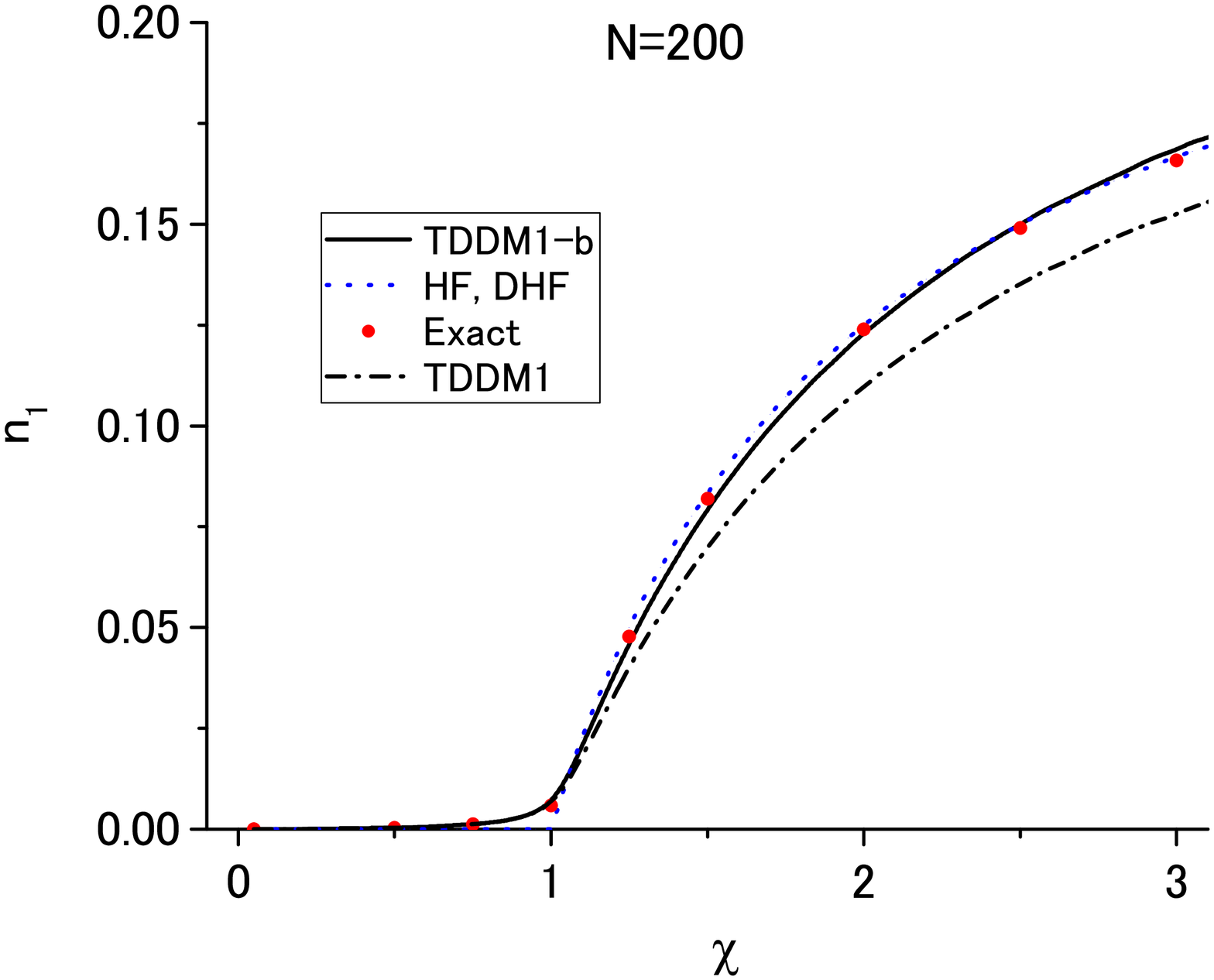}
}
\caption{Same as fig. \ref{20n3l} but for $N=200$.}
\label{200n3l}
\end{figure}

\begin{figure}
\resizebox{0.5\textwidth}{!}{
\includegraphics{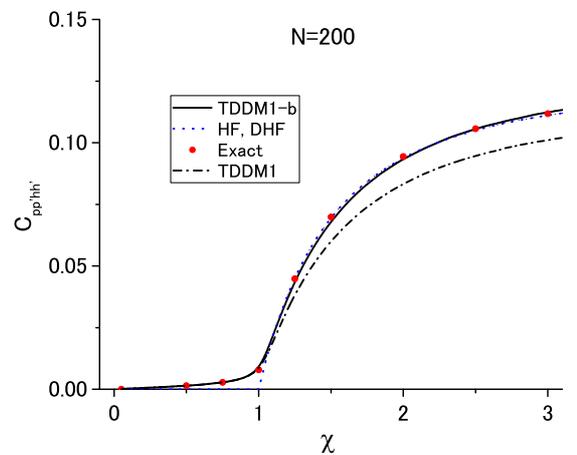}
}
\caption{Same as fig. \ref{20c3l} but for $N=200$.}
\label{200c3l}
\end{figure}

The ground-state energy calculated in TDDM1-b (solid line) is shown in figs. \ref{20e3l} and \ref{200e3l} as a function of $\chi$
for $N=20$ and $200$, respectively. The exact values are given by the dots. The dashed and dot-dashed lines depict the results in TDDM and TDDM1, respectively.
The results in HF and DHF($\chi>1$) are shown with the dotted line but in the case of $N=200$ they cannot be easily distinguished from the TDDM1-b results in the scale of the figure.
The results in TDDM2 are given with the two-dot-dashed line in fig. \ref{20e3l}. In the case of $N=200$ they are close to those in TDDM as in the case of the two-level Lipkin model
and are not shown in fig. \ref{200e3l}.
In contrast to the case of the two-level Lipkin model TDDM strongly overestimates two-body correlations in the three-level Lipkin model even for $N=200$.  
This is due to the fact that the three-body correlation matrix cannot be neglected in the three-level Lipkin model as discussed above.
DHF gives a good description of the ground-state energy for $N=200$ as in the case of the two-level Lipkin model. The results in
TDDM1-b agree well with the exact solutions both in $N=20$ and $200$. TDDM1 underestimates the correlation effects for large $\chi$ because the phh--phh
component of the three-body correlation matrix given by eq. (\ref{purt2}) is non-vanishing.
The occupation probability $n_1$ of the single-particle level labeled 1 and the 2p--2h element of the two-body correlation matrix calculated in TDDM1-b are also in good agreement with the exact values as 
shown in figs. \ref{20n3l}--\ref{200c3l}. We have checked that the agreement of the TDDM1-b results with the exact solutions extends at least to $\chi=10$ in the case of $N=200$.
The results in TDDM1 are not so good as those in TDDM1-b but the agreement with the exact values are reasonable in the transition region ($\chi\approx1$).
In the case of the two-level Lipkin model the ground state energies calculated in TDDM1-b approximately come in the middle between the results in TDDM and TDDM1.
 
\begin{figure}
\resizebox{0.5\textwidth}{!}{
\includegraphics{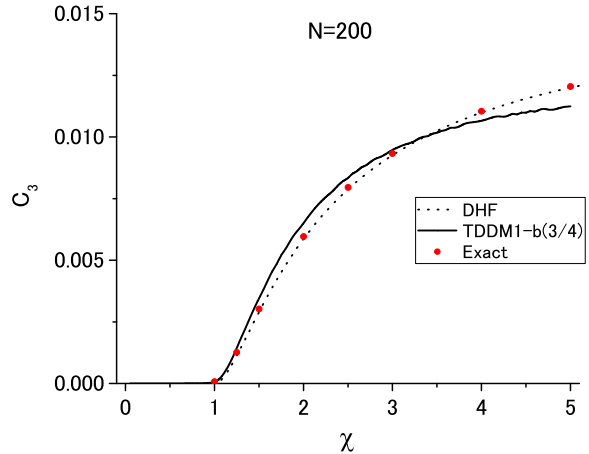}
}
\caption{Three-body correlation matrix $C_{11,02,13,11,12,03}~(C_3)$ used in TDDM1-b (solid line) multiplied by a factor $3/4$
as a function of $\chi$ for $N=200$. 
The results in HF and DHF($\chi>1$) are shown with the dotted line. The exact values are given by the dots.}
\label{200c33l}
\end{figure}

In the following we try to explain why TDDM1-b gives good results for the three-level Lipkin model.
In the large $N$ case of the three-level Lipkin model (and also the two-level Lipkin model) the coupling of $C_{\rm pp'hh'}$ and $C_{\rm hh'pp'}$
to $C_{\rm ph'p'h}$ described by the $H_{\alpha\beta\alpha'\beta'}$ term in eq. (\ref{r2}) is dominant (see eq. (\ref{h-term})).
The three-body correlation matrix of the form 
$C_{\rm pp'p''hp'h''}$,
which enters the equation of motion for $C_{\rm ph'p'h}$, is neglected in TDDM1-b(and also in TDDM1). This is because such an element of the three-body correlation matrix
is of higher order than eq. (\ref{purt1}) in the perturbative regime. In the strongly interacting region of the three-level Lipkin model $C_{\rm pp'p''hp'h''}$ becomes non-vanishing as discussed above. 
As a consequence of the neglect of $C_{\rm pp'p''hp'h''}$, $C_{\rm ph'p'h}$ is overestimated in TDDM1-b. 
If $C_{\rm pp'p''hp'h''}$ is assumed to be $n_{\rm p'p'}C_{\rm pp''hh''}/2$ using eqs. (\ref{npp}) and (\ref{pphh}), the overestimation in eq. (\ref{r2}) for $C_{\rm ph'p'h}$ can be evaluated to be 
$n_{1}C_{\rm pp'hh'}/(n_0-2n_1)$, where $n_0$ is the occupation probability of the single-particle level labeled $0$.
It is impossible to analytically calculate its contribution to
$C_{\rm ph'p'h}$ but comparison of $C_{\rm ph'p'h}$ in TDDM1-b and that in DHF
which is given by eq. (\ref{cphph}) shows that the deviation of $C_{\rm ph'p'h}$
is well expressed as 
\begin{eqnarray}
\frac{C_{\rm pp'hh'}^2}{2(n_0-n_1)}.
\label{over}
\end{eqnarray}
The quadratic form is understandable because $n_1$ is determined by $C_{\rm pp'hh'}$ in eq. (\ref{n1}).
In the equation of motion for $C_{\rm pp'hh'}$ in TDDM1-b one of the four terms in eq. (\ref{r2}) which include the three-body correlation matrix ($C_3$) cancels the overestimated part of $C_{\rm ph'p'h}$:
eq. (\ref{over}) is multiplied by $-2(n_0-n_1)$ due to the occupation factors in eq. (\ref{h-term}).
Consequently,
the three remaining $C_3$ terms in TDDM1-b play a role as the effective $C_3$ terms.
In fig. \ref{200c33l} the three quarters of eq. (\ref{purt1}) calculated in TDDM1-b (solid line) are compared with the exact values of $C_3$ (dots) 
and the DHF values (dotted line) given by eq. (\ref{c3dhf}). The TDDM1-b values well simulates the $\chi$ dependence of the exact $C_3$.
This is the reason behind why the simple truncation scheme eq. (\ref{purt1}) gives good results.

Our study of the three-level Lipkin model has shown that contrary to the two-level 
case the three-body correlation matrix cannot be neglected in the large $N$ (thermodynamic) 
limit. Working in the symmetry unbroken basis, the cumulant approximation to 
the correlated part of the three body density matrix works very well for the 
ground state energy if only the pph--pph term is retained. We gave arguments why 
the phh--phh particle contribution must be discarded. 
However, the validity of TDDM1-b remains 
restricted to the particularities of the model and conclusions for the general 
case cannot be made.
On the other hand, staying 
in the symmetry unbroken description, one cannot describe the rotational 
spectrum which emerges in the three-level Lipkin model \cite{li} if the two upper levels 
become degenerate as is the case of eq. (\ref{3lH}).
For such cases with a spontaneously broken 
symmetry it is in general better to start with a symmetry broken basis and to 
recover good symmetry with projection techniques. 
DTDDM which consists of the simplest truncation scheme for $C_3$ and the symmetry broken single-particle basis
may be applicable for such cases.
In the past we applied our cumulant approximation eqs. (\ref{purt1}) and (\ref{purt2}) to the more realistic 
case of $^{16}$O with promising results \cite{toh2015}. In these 
examples the symmetry unbroken (spherical) description is certainly the choice 
to be used.

\begin{figure}
\resizebox{0.5\textwidth}{!}{
\includegraphics{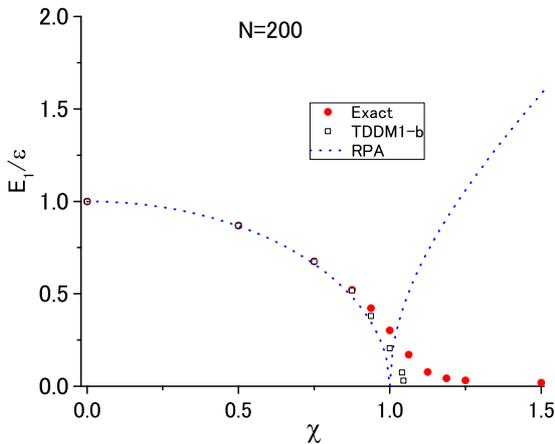}
}
\caption{Excitation energy of the first excited state in TDDM1-b (open squares) as a function of $\chi$ for $N=200$. The results in
RPA are shown with the dotted line and the exact values are given by the dots.}
\label{200e13l}
\end{figure}

\begin{figure}
\resizebox{0.5\textwidth}{!}{
\includegraphics{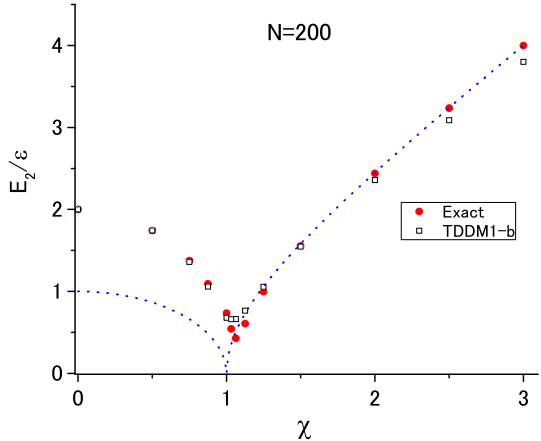}
}
\caption{Excitation energy of the second excited state in TDDM1-b (open squares) as a function of $\chi$ for $N=200$. The results in
RPA for the first excited states are shown with the dotted line. The dots depict
the exact values for the lowest excited state with $\langle L_0^2\rangle=0$.}
\label{200e23l}
\end{figure}

The excited states of the three-level Lipkin model can be calculated in the same manners as those used for the two-level Lipkin model. 
The results obtained from the TDDM1-b based approaches are shown in figs. \ref{200e13l} and \ref{200e23l}.
The exact eigenstates are labeled by the quantum number $\langle L_0^2\rangle=0,~1,~2,\cdot\cdot\cdot$ where $L_0=i(K_{12}-K_{21})$ \cite{delion}
with $K_{12}(=K_{21}^\dag)=\sum_i c^\dag_{1i}c_{2i}$, and the ground state has $\langle L_0^2\rangle=0$.
In fig. \ref{200e23l} the exact values for the lowest excited state with the same quantum number as the ground state are shown.
Both the first and second excited states in TDDM1-b have $\chi$ dependence which is similar to the case of the two-level Lipkin model. The first excited state in TDDM1-b collapses beyond $\chi=1.045$. 
A more accurate treatment of the three-body correlation matrix than TDDM1-b is required. 

\section{Summary}
We studied the time-dependent density-matrix theory ( TDDM) in the large $N$ and strong coupling limits of the Lipkin models.
It was found that in the two-level Lipkin model the ground state calculated using the original truncation scheme of TDDM 
where the three-body correlation matrix is completely neglected
approaches the exact solution with increasing number of particles. It was pointed out that this is related to the fact that the deformed Hartree-Fock approximation (DHF)
gives the exact ground-state energies in the large $N$ limit. The relation between the occupation matrix in DHF and the correlation matrix in TDDM was also discussed.
The small amplitude limit of TDDM was also found to describe the excited states of the two-level Lipkin model
very well in the 'spherical' region. In the 'deformed' region the first excited state falls to zero as it should. 
However, it falls to zero with a finite slope contrary to the exact solution which goes to zero asymptotically.
It indicates that the three-body correlation matrix neglected in TDDM still plays some role in excited states. 

It was shown that in the three-level Lipkin model a simple truncation scheme where the three-body correlation matrix 
is approximated by the square of the two-body correlation matrix gives good results for the ground-state properties. 
Also the first two excited states are quite well reproduced far into the strongly correlated (deformed) region working in the symmetry unbroken basis.

We pointed to the fact that the influence of the three-body correlation matrix in the two- and three-level Lipkin models is quite different. 
While in the former model the genuine three-body correlations totally disappear in the large $N$ limit, they partially vanish in the latter model. 
We gave reasons for this different behavior and suggested that for other spin models similar studies could eventually also give insight into the role of three-body correlations there.

The fact that three-body correlation matrix in the strong coupling ('deformed') regions 
is apparently quite model dependent is perturbing and sheds some doubt of the 'blind' applicability 
of the cumulant approximation (TDDM1) to the three-body correlation matrix in 'deformed' regions while working in the spherical basis.
On the other hand more realistic applications to $^{16}$O \cite{toh2015} and other models \cite{ts2014} 
where the three-body correlation matrix does not disappear have shown in the past 
that the cumulant approximation is quite powerful for symmetry unbroken systems. It was also shown that TDDM1 gives reasonable results for the three-level Lipkin model
as long as the interaction is not so strong. 
Therefore we think that the cumulant approximation is applicable to realistic systems.

\appendix
\section{Terms in eq. (\ref{r2})}
 
The terms in eq. (\ref{r2}) are given below.
$B_{\alpha\beta\alpha'\beta'}$ describes the 2p--2h and 2h--2p excitations. 
\begin{eqnarray}
B_{\alpha\beta\alpha'\beta'}&=&\sum_{\lambda_1\lambda_2\lambda_3\lambda_4}
\langle\lambda_1\lambda_2|v|\lambda_3\lambda_4\rangle_A
\nonumber \\ 
&\times&
[(\delta_{\alpha\lambda_1}-n_{\alpha\lambda_1})(\delta_{\beta\lambda_2}-n_{\beta\lambda_2})
n_{\lambda_3\alpha'}n_{\lambda_4\beta'}
\nonumber \\
&-&n_{\alpha\lambda_1}n_{\beta\lambda_2}(\delta_{\lambda_3\alpha'}-n_{\lambda_3\alpha'})
(\delta_{\lambda_4\beta'}-n_{\lambda_4\beta'})].
\nonumber \\
\end{eqnarray}
Particle -- particle and h--h correlations 
are described by $P_{\alpha\beta\alpha'\beta'}$
\begin{eqnarray}
P_{\alpha\beta\alpha'\beta'}&=&\sum_{\lambda_1\lambda_2\lambda_3\lambda_4}
\langle\lambda_1\lambda_2|v|\lambda_3\lambda_4\rangle
\nonumber \\ 
&\times&
[(\delta_{\alpha\lambda_1}\delta_{\beta\lambda_2}
-\delta_{\alpha\lambda_1}n_{\beta\lambda_2}
-n_{\alpha\lambda_1}\delta_{\beta\lambda_2})
{ C}_{\lambda_3\lambda_4\alpha'\beta'}
\nonumber \\
&-&(\delta_{\lambda_3\alpha'}\delta_{\lambda_4\beta'}
-\delta_{\lambda_3\alpha'}n_{\lambda_4\beta'}
-n_{\lambda_3\alpha'}\delta_{\lambda_4\beta'})
{ C}_{\alpha\beta\lambda_1\lambda_2}].
\nonumber \\
\end{eqnarray}
$H_{\alpha\beta\alpha'\beta'}$ describes p--h correlations.
\begin{eqnarray}
H_{\alpha\beta\alpha'\beta'}&=&\sum_{\lambda_1\lambda_2\lambda_3\lambda_4}
\langle\lambda_1\lambda_2|v|\lambda_3\lambda_4\rangle_A
\nonumber \\ 
&\times&
[\delta_{\alpha\lambda_1}(n_{\lambda_3\alpha'}{ C}_{\lambda_4\beta\lambda_2\beta'}
-n_{\lambda_3\beta'}{ C}_{\lambda_4\beta\lambda_2\alpha'})
\nonumber \\
&+&\delta_{\beta\lambda_2}(n_{\lambda_4\beta'}{ C}_{\lambda_3\alpha\lambda_1\alpha'}
-n_{\lambda_4\alpha'}{ C}_{\lambda_3\alpha\lambda_1\beta'})
\nonumber \\
&-&
-\delta_{\alpha'\lambda_3}(n_{\alpha\lambda_1}{ C}_{\lambda_4\beta\lambda_2\beta'}
-n_{\beta\lambda_1}{ C}_{\lambda_4\alpha\lambda_2\beta'})
\nonumber \\
&-&\delta_{\beta'\lambda_4}(n_{\beta\lambda_2}{C}_{\lambda_3\alpha\lambda_1\alpha'}
-n_{\alpha\lambda_2}{ C}_{\lambda_3\beta\lambda_1\alpha'})].
\nonumber \\
\label{h-term}
\end{eqnarray}

\end{document}